\documentclass[preprints,article,accept,moreauthors,pdftex]{Definitions/mdpi} 

\firstpage{1} 
\pubvolume{1}
\issuenum{1}
\articlenumber{0}
\pubyear{2021}
\copyrightyear{2020}
\datereceived{} 
\dateaccepted{} 
\datepublished{} 

\usepackage[maxfloats=256]{morefloats}
\Title{Applying the horizontal visibility graph method to study irreversibility of electromagnetic turbulence in non-thermal plasmas}

\TitleCitation{Horizontal visibility graph method to study irreversibility of electromagnetic turbulence in plasmas} 

\Author{Belén Acosta $^{1,*,\dagger}$\orcidA{}, Denisse Pastén$^{1,*,\dagger}$\orcidB{} and Pablo S. Moya$^{1,*,\dagger}$\orcidC{}}

\AuthorNames{Belén Acosta, Denisse Pastén, Pablo S. Moya}

\AuthorCitation{Acosta, B.; Pastén, D.; Moya, P. S.}

\address[1]{%
$^{1}$ \quad Departamento de Física, Facultad de Ciencias, Universidad de Chile, Santiago, Chile}

\corres{Correspondence: bacostaazocar@gmail.com (B.A.); denisse.pasten.g@gmail.com (D.P.); pablo.moya@uchile.cl (P. S.M.);}
\firstnote{All authors contributed equally to this work.} 

\abstract{ One of the fundamental open questions in plasma physics is the role of non-thermal particles distributions in poorly collisional plasma environments, a system commonly found throughout the Universe, e.g. the solar wind and the Earth’s magnetosphere correspond to natural plasma physics laboratories in which turbulent phenomena can be studied. Our study perspective is born from the method of Horizontal Visibility Graph (HVG) that has been developed in the last years to analyze time series avoiding the tedium and the high computational cost that other methods offer. Here we build a complex network based on directed HVG technique applied to magnetic field fluctuations time series obtained from Particle In Cell (PIC) simulations of a magnetized collisionless plasma to distinguish the degree distributions and calculate the Kullback-Leibler Divergence (KLD) as a measure of relative entropy of data sets produced by processes that are not in equilibrium. First, we analyze the connectivity probability distribution for the undirected version of HVG finding how the Kappa distribution for low values of $\kappa$ tends to be an uncorrelated time series, while the Maxwell-Boltzmann distribution shows a correlated stochastic processes behavior. Then, we investigate the degree of temporary irreversibility of magnetic fluctuations self-generated by the plasma, comparing the case of a thermal plasma (described by a Maxwell-Botzmann velocity distribution function) with non-thermal Kappa distributions. We have shown that the KLD associated to the HVG is able to distinguish the level of reversibility associated to the thermal equilibrium in the plasma because the dissipative degree of the system increases as the value of $\kappa$ parameter decreases and the distribution function departs from the Maxwell-Boltzmann equilibrium.}

\begin{document}

\section{Introduction}

In a turbulent collisionless plasma (in which Coulomb collisions are neglected), movement on a kinetic scale (spatial scales of the order of the particles Larmor radius or skin-depth) occurs in a chaotic manner, and is determined by large-scale collective behavior and also localized small-scale processes. This kind of system can be commonly found throughout the Universe. The solar wind and the Earth's magnetosphere correspond to natural plasma physics laboratories in which plasma phenomena can be studied~\cite{KamideChian}. Some non-linear phenomena include magnetic reconnection~\cite{Yamada2010}, collisionless shocks~\cite{BaloghTreumann}, electromagnetic turbulence~\cite{brunocarbone2013}, collisionless wave-particle interactions~\cite{Yoon2017}, or plasma energization and heating~\cite{marsch2006}. One of the fundamental open questions in plasma physics is the understanding of the energy equipartition between plasma and the electromagnetic turbulence, and the role of non-thermal plasma particles distributions ubiquitous in poorly collisional plasma environments.

One of the most used approaches to model non-thermal plasma systems is through the representation of the plasma velocity distribution function (VDF) using the well-known Tsallis or Kappa distributions. First proposed by Olbert~\cite{Olbert1968} and Vasyliunas~\cite{Vasyliunas1968} to fit electron measurements in the magnetosphere, it is accepted that Kappa distributions are the most common state of electrons~\cite[see e.g.][]{maksimovic1997kinetic,Pierrard2017}, and have been observed in space in the solar wind~\cite{Livadiotis2018,Lazar2020}, the Earth's magnetosphere~\cite{Espinoza2018,Eyelade2021} or other planetary environments~\cite{Dialynas2017}. These distributions resolve both, the quasi-thermal core and the power-law high energy tails measured by the $\kappa$ parameter, and correspond to a generalization of the Maxwell-Boltzmann distribution, achieved when $\kappa \to \infty$.   
Kappa distributions have been widely studied in the framework of non-equilibrium statistical mechanism as corresponding to a class of expected probability distribution function when the system exhibits non-extensive entropy~\cite{Tsallis1988,Tsallis2009,Yoon2019}. Regarding plasma physics, it has been found that in Kappa-distributed plasmas, the non-thermal shape of the distribution function plays a key role on the details of kinetic processes such as wave-particle interactions~\cite{Lazar2016,vinas2017}, that mediate the collisionless relaxation of unstable plasma populations~\cite{Lazar2019,Lopez2019,moya2020}. Moreover, even in the absence of instabilities, in a plasma with finite temperature the random motion of the charged particles composing the plasma produces a finite level of electromagnetic fluctuations. These fluctuations, known as quasi-thermal noise, can be explained by a generalization of the Fluctuation-Dissipation Theorem~\cite[see e.g.][and references therein]{Navarro2014,vinas2015}, and have been studied in the case of thermal and non-thermal plasma systems. Recent results have shown that the fluctuations level in plasmas including supra-thermal particles following a Kappa distribution is enhanced with respect to plasma systems in thermodynamic equilibrium~\cite{vinas2014role,vinas2015,Lazar2018b}. 

Regardless the nature of the distribution function (thermal or non-thermal), plasmas show  a self-organized critical behavior~\cite{sharma2016SOC} allowing the introduction of concepts from complex systems to study this criticality. That methods are applied both in data sets and models~\cite{chapman2018NPG,dominguez2018pop,dominguez2020npg,munoz2018npg}. Some authors have suggested that the change of fractals and multifractals indexes could be associated with dissipative events or related to the solar cycle, proposing a relation between multifractality and physical processes in plasmas, among them solar cycle, Sun-Earth system or theoretical models of plasmas~\cite{wawrzaszek2019astrojournal,chapman2018NPG}. Another studies show a relation between intermittency fluctuations and multifractal behavior while the fluctuations at kinetic-scales reveal a monofractal behavior~\cite{alberti2019multifractal,owen2020multifractal, Chhiber2021}.  \citet{wawrzaszek2019astrojournal} apply a multifractal formalism to the solar wind suggesting a relation between the intermittency and the degree of multifractality.  
Those studies show different time series analysis in plasmas. But not only fractals and multifractals could be useful in the study of time series, complex networks, particularly the Visibility Graph method allow a simple and direct time series analysis in self-organized critical phenomena, such as earthquakes~\cite{luctelesca2020time}, macroeconomic systems~\cite{na2012} or biological systems~\cite{zheng2021}. 

The method of Visibility Graph~\cite{lacasa2009time} has been developed in the last years, it allows us study and analyze time series avoiding the tedium and the high computational cost that other methods offer. The visibility algorithm proceeds to map a times series into a complex network under a geometric principle of visibility, in this sense, the algorithm could be considered as a geometric transform of the time series in which this method decomposes a time series in connections between nodes that could be repeated or not, forming a particular weave that represents the time series as a geometric object. Inside the Visibility Graph algorithm, we can use a simplification of it, the Horizontal Visibility Graph (HVG), where the nodes are connected if it is possible to draw a horizontal line between two nodes.The HVG has been applied to different systems, from earthquakes~\cite{luctelesca2020time} and plasmas~\cite{suyal2014visibility} to chaotic processes~\cite{lacasa2010time}. 
The visibility graph is constructed under the visibility criterion, two data $(t_a, y_a)$ and $(t_b, y_b)$ in the time series look at each 
other if there is a data $(t_c, y_c)$, with $t_a < t_c < t_b$ such that satisfy the condition~\cite{lacasa2009time,lacasa2012time}:

     \begin{equation}
     y_c < y_a + (y_b - y_a)\frac{t_c - t_a}{t_b - t_a} \,.   
     \end{equation}

In the field of space plasma physics, the VG has been applied to solar flares~\cite{gheibi2017astrojournal,najafi2020} and solar wind measurements~\cite{suyal2014visibility}. In particular,~\citet{najafi2020} show a complete and detailed analysis of solar flares through a combination of two methods of complex networks: a time-based complex network supported by the work of~\citet{abe2006} and the VG method proposed by~\citet{telesca2012}. They characterize solar flares based in the probability distribution of connectivity and clustering coefficient finding a good agreement with results obtained in another works with seismic data sets. In addition,~\citet{suyal2014visibility} studied the irreversibility of velocity fluctuations. Through the HVG method they calculated Kullback Leibler Divergence (KLD) of the fluctuations, and found that irreversibility in solar wind velocity fluctuations show a similar behavior at different distances from the Sun, and that there is a dependence of the KLD with the solar cycle. The KLD or relative entropy value, is a measure of temporary irreversibility of data sets produced by processes that are not in equilibrium, and gives information on the production of entropy generated by the physical system, considering that a high degree of irreversibility as a chaotic and dissipative system~\cite{lacasa2009time}. 
Under this context, recent results by~\citet{Acosta2019} have suggested that the use of the HVG method can provide valuable information to characterize turbulence in collisionless plasmas, and that the KLD may be used as a proxy to establish how thermal or non-thermal are the velocity distributions of a plasma, only by looking at the magnetic fluctuations and their properties.

Here we build a complex network based on the HVG technique~\cite{lacasa2012time} applied to magnetic field fluctuations time series obtained from Particle In Cell (PIC) simulations of a magnetized collisionless plasma. We analyze the degree of irreversibility of magnetic fluctuations self-generated by the plasma, comparing the case of a thermal plasma (described by a Maxwell-Botzmann VDF) with the fluctuations generated by non-thermal Kappa distributions. In order to understand the degree of the irreversibility as a parameter that could be related to the shape of the particles velocity distributions, we computed the KLD for different values of the $\kappa$ parameter for comparative purposes and analyzed their time evolution throughout each simulation. The paper is organized as follows. In Section~\ref{HVG} the methods and techniques are described, Section~\ref{PIC} shows the model used to build the time series, and in Section~\ref{Results} the results are presented. Finally, in section~\ref{Discussion} we discuss our results and present the conclusions of our study.

\section{Horizontal Visibility Graph: Mapping time series to network} 
\label{HVG}

\begin{figure}[ht]
\centering
\includegraphics[width=0.8\linewidth]{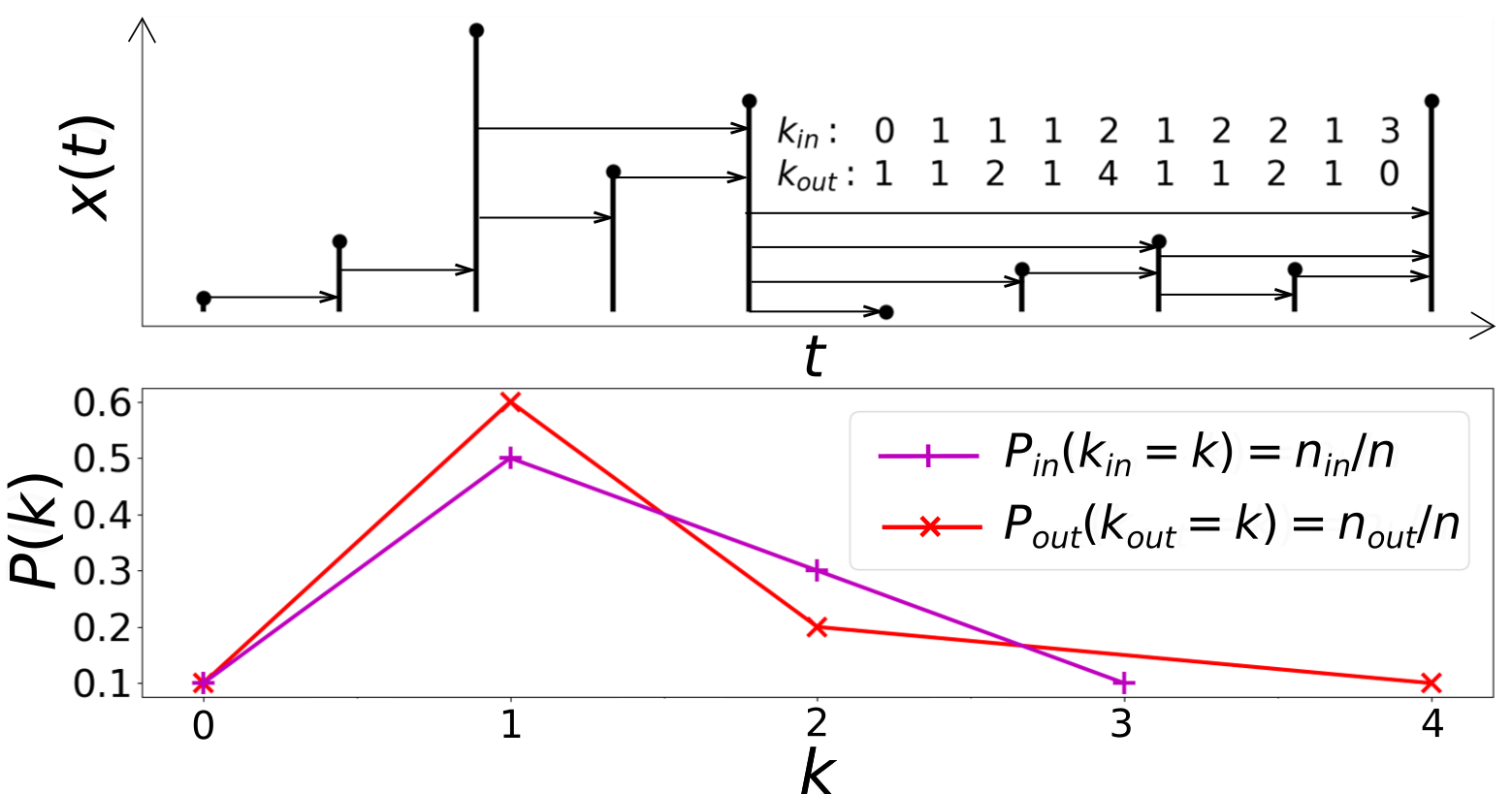}
\caption{Construction of Horizontal Visibility Graph. Top, a time series where the degree $k_{\text{in}}$ for in-going links and $k_{\text{out}}$ for out-going links of each of the $n=10$ nodes are detailed. Bottom, probability distribution $P$ in relation to degree $k$, where $n_{\text{in}}$ and $ n_{\text{out}}$ correspond to the frequency of appearance of the degrees $k_{\text{in}} $ and $k_{\text{out}} $ respectively, defining $P_{\text{in}} $ and $P_{\text{out}}$.}
  \label{fig:HVG}
\end{figure}

We use the directed version of the Horizontal Visibility Graph, method that models the time series as a directed network according to a geometric criteria that considers the magnitude of each data and then evaluates its horizontal visibility with the other data in the series in the direction of time (see Figure ~\ref{fig:HVG} for a graphical illustration). More precisely, first let $\left\{x_i \right\}_{i=1,. . .,n}$ be a time series of $n$ data. The algorithm consists of assigning each data of the series to a node. Then, two nodes $i$ and $j$ in the graph are connected if one can draw a horizontal line in the time series joining $x_i$ and $x_j$ that does not intersect any intermediate data height. Hence, $i$ and $j$ are two connected nodes if the following geometrical criterion is fulfilled within the time series~\cite{luque2009horizontal}: 

    \begin{equation}
    x_i,\; x_j > x_m \quad \text{ for all } m \text{ such that } \quad i < m < j \,.
    \end{equation}

Then, for a graph directed in the direction of the time axis, for a given node two different degrees are distinguished. These are the in-going degree $k_{\text{in}}$, related to how many nodes see a given node $i$, and an out-going degree $k_{\text{out}}$ that is the number of nodes that node $i$ sees~\cite{luque2009horizontal}. With this temporal direction, causality in the network is implicit in each degree, since the input degree $k_{\text{in}}$ is associated with links of a node with other nodes of the past. Meanwhile, the degree of output $k_{\text{out}}$ is associated with the links with nodes of the future~\cite{lacasa2012time}. From the properties of these connections, it can be said that if the graph remains invariant under the reversion of time, it could be stated that the process that generated the series is conservative~\cite{luque2009horizontal}.

To further detail the dynamics between nodes, degree distributions play a fundamental role. The degree distribution of a graph describes the probability of an arbitrary node to have degree $k$ (i.e. $k$ links)~\cite{newman2003properties}, and it becomes absolutely necessary to measure the difference between $P_{\text{in}}$ and $P_{\text{out}}$ to understand the incidence of the action of time on the process that originates the series (see Figure~\ref{fig:HVG} for illustration of both degree distributions).

\subsection{Kullback-Leibler Divergence: Measuring irreversibility}
\label{KLD}

We are interested in measuring the irreversibility of the time series since it is indicative of the presence of nonlinearities in the underlying dynamics and is associated with systems driven out-of-equilibrium~\cite{kawai2007dissipation, parrondo2009entropy}. A stationary process $X(t)$ is said to be statically reversible if for each $N$ the series  $\{X(t_1), ..., X(t_N)\}$ y $\{X(t_N), ..., X(t_1)\}$ have the same probability of degree distribution~\cite{weiss1975time}. In other words, we speak of reversibility if for all the data of the series developed in its natural time and its inverse time, the same degree distribution is presented in each case. A method was proposed to measure real-valued time series irreversibility which combines two different tools: the HVG algorithm and the Kullback-Leibler Divergence~\cite{lacasa2012time}. The degree of irreversibility of the series is then estimated by the Kullback-Leibler Divergence, a way to evaluate the difference between $P_{\text{in}}$ and $P_{\text{out}}$ of the associated graph. The KLD is defined by

    \begin{equation}
    D[P_{\text{out}}(k)||P_{\text{in}}(k)] = \sum _k P_{\text{out}}(k)\log \frac{P_{\text{out}}(k)}{P_{\text{in}}(k)}\,.
    \label{eq:KLD}
    \end{equation}

In the above definition, we use the convention that $0\log \frac{0}{0} = 0$, $0\log \frac{0}{q} = 0$ and $p\log \frac{p}{0} = \infty$, based on continuity arguments. Thus, if there is any symbol $x \in X$ such that $p(x) > 0$ and $q(x) = 0$, then $D(p||q) = \infty$ ~\cite{cover2006elements}. The KLD is always non-negative and is zero if and only if $P_{\text{out}} = P_{\text{in}}$, so if $D \rightarrow 0$, the system has a low degree of irreversibility. We estimate the relative entropy production of the physical process that generated the data, since this measure gives lower bounds to the entropy production~\cite{lacasa2012time}.

\section{Particle In Cell Simulations: Thermal and non-thermal plasma particle distributions}
\label{PIC}

To build time series of magnetic fluctuations produced by a collisionless plasma we performed PIC simulations. The simulations treat positive ions and electrons as individual particles that are self-consistently accelerated by the electric and magnetic field through the charge and current densities collectively produced by themselves. For our study we consider a so-called 1.5D PIC code, that resolves the movement of the particles in one dimension but computes the three components of the velocity of each particles, and therefore the three components of the current density. Our code has been tested and validated in several studies ~\cite[see e.g.][]{Lopez2017,Lazar2018}, and technical details about the used numerical schemes can be found in~\cite{vinas2014role}.

We simulate a magnetized plasma composed by electrons and protons with masses $m_e$ and $M_p$, respectively, and realistic mass ratio $M_p/m_e \sim 1836$. We assume the warm plasma as quasineutral, in which both species have number density $n_0$, such that $\omega_{pe}/\Omega_e = 5$. Also, $\omega_{pe} = \left(4\pi n_o e^2/m_e\right)^{1/2}$ is the plasma frequency, $\Omega_e = (eB_0)/(m_e c)$ is the electron gyro-frequency, $e$ is the elementary charge, $c$ the speed of light, and $B_0$ is the background magnetic field. Our code solve the equations in a one dimensional grid with periodic boundary conditions, and the background magnetic field aligned with the spatial grid ($\mathbf{B_0} = B_0 \hat x$). To resolve the kinetic physics of electrons we set up a grid of length $L = 256\,\lambda_e$, where $\lambda_e = \omega_{pe}/c$ is the electron inertial length. We divide the grid in $N=2048$ cells, initially with 1000 particles per species per cell, and run the simulation up to $t = 1330.72/\Omega_e$ in time steps of length $dt = 0.01\Omega_e$. For each simulation we initialize the particles velocities following an isotropic VDF $f_j(v)$, with $j =e $ for electrons and $j=p$ for protons, and $v$ represents the velocity. For the case of a plasma in thermodynamic equilibrium $f_j$ corresponds to a Maxwell-Boltzmann distribution
\begin{equation}
\label{eq:fmaxwell}
    f_j(v) = \frac{n_0}{\pi^{3/2}\alpha^3_j}\,\exp\left(-\dfrac{v^2}{\alpha^2_j}\right)\,,
\end{equation}
and in the case of a non-thermal plasma $f_j$ is given by a Kappa distribution. Namely:
\begin{equation}
\label{eq:fkappa}
    f_j(v) = \frac{n_0}{\pi^{3/2}\alpha^3_j}\,\frac{\Gamma(\kappa_j)}{\kappa_j \Gamma(\kappa_j -1/2)}\,\left(1+\dfrac{1}{\kappa_j}\,\dfrac{v^2}{\alpha^2_j}\right)^{-(\kappa_j+1)}\,.
\end{equation}
Here, $\alpha_j = \left(2 k_B T_j/m_j\right)^{1/2}$ is the thermal velocity of the distribution, $\kappa_j$ and $T_j$ are the kappa parameter and the temperature of each species, and $k_B$ is the Boltzmann constant. Also, $\Gamma$ corresponds to the Gamma function, and note that Kappa distributions [Equation~\eqref{eq:fkappa}] becomes the Maxwell-Boltzmann distribution [Equation~\eqref{eq:fmaxwell}] in the limit $\kappa\to\infty$. However, for kappa values $\kappa\gtrapprox 10$ the Kappa and Maxwellian VDFs are relatively similar.  

Following all these consideration, for our study we run and compare the results of three different simulations with different values of the electron $\kappa_e$ parameter. Case 1: a plasma in thermal equilibrium with electrons following a Maxwell-Boltzmann distribution given by Equation~\eqref{eq:fmaxwell}; case 2: non-thermal electrons following Equation~\eqref{eq:fkappa} with $\kappa_e=3$, representing a system far from thermodynamic equilibrium; and case 3: a plasma with $\kappa_e=15$, also non-thermal but closer to equilibrium. In addition, to isolate the effects of thermal or non-thermal electrons, for all three cases we consider protons following a Maxwellian; i.e. $\kappa_p\to\infty$. Finally, for all cases, we consider a plasma with temperature $T_j$, such that the plasma beta parameter is $\beta_j = 8\pi n_0 k_B T_j/B^2_0 = 0.01$ for both species; i.e. $\beta_e=\beta_p=0.01$.
\begin{figure}[ht]	
\widefigure
\includegraphics[width=\linewidth]{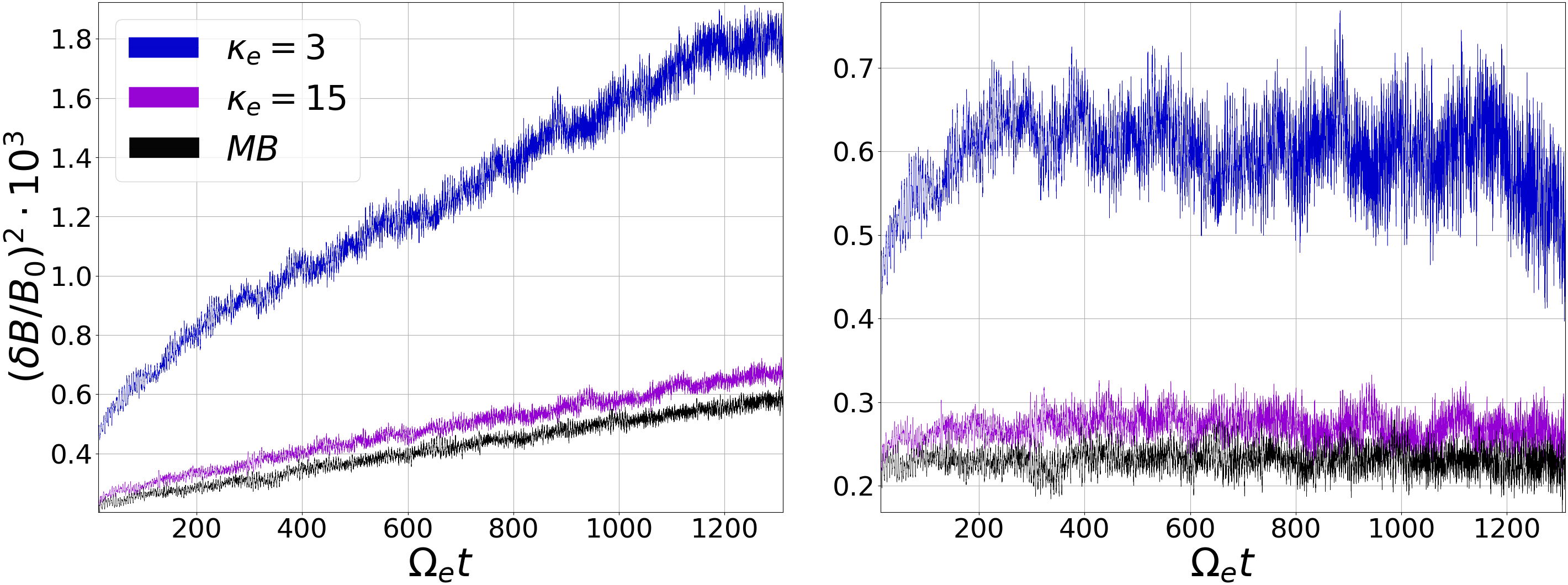}
\caption{(Left) Average magnetic field energy density fluctuations $(\delta B/B_0)^2$ as a function of time obtained from PIC simulations for Maxwell-Boltzmann (where MB represents $\kappa_e\to\infty$) and Kappa distributions considering different values of the $\kappa_e$ parameter. (Right) Detrended average magnetic field energy density magnitude.}
\label{fig:PIC}
\end{figure} 

As already mentioned, even though a collisionless isotropic plasma is a system at equilibrium according to the Vlasov Equation, the plasma will develop a certain level of magnetic fluctuations spontaneously produced by the motion of the charged particles~\cite{Navarro2014,vinas2014role,vinas2015,Lazar2018b}. This is precisely the situation of our study for any of the three simulations (three cases) we have performed. Figure~\ref{fig:PIC} shows the average magnetic field energy density fluctuations $(\delta B/B_0)^2$ as a function of time, for all three cases. To build these time series, at each time step we have computed the transverse magnetic fluctuations at the plane perpendicular to $\mathbf{B_0}$ and have averaged the magnitude of the fluctuations at each grid point.
Figure~\ref{fig:PIC} (left) shows the fluctuations time series for $\kappa_e=3$ (blue), $\kappa_e=15$ (purple), and the $\kappa_e\to\infty$ or Maxwell-Boltzmann distribution (black). As expected, we can see that the level of fluctuations increases with decreasing value of $\kappa_e$, and that the behavior of the fluctuations with $\kappa_e=15$ is fairly similar to the Maxwellian case. In addition, Figure~\ref{fig:PIC} (right) presents the time series of the detrended fluctuations, where we can see that the amplitude of the fluctuations also increases as $\kappa_e$ decreases. In the next section the HVG method will applied to all these time series. 

\section{Results}
\label{Results}

We apply the HVG method to study time series of magnetic fluctuations obtained from the PIC simulations. Considering the Maxwellian and Kappa distributions we follow the HVG algorithm and build complex networks for three cases: Maxwellian distribution (thermal equilibrium with electrons), $\kappa_e =$ 3 (non-thermal electrons), $\kappa_e =$ 15 (non-thermal electrons, but closer to the equilibrium), using the original and the detrended time series (see Figure~\ref{fig:PIC}). First, we build the complex network and we calculate the in-going and out-going degree distribution for each time series, in order to characterize this distribution for each case.
In this sense, the probability distribution of the degree gives information related to the correlations in a process, in this case, time correlations. According to the theorem for uncorrelated time series~\cite{luque2009horizontal},  the degree distribution of the horizontal visibility graph associated with a bi-infinite sequence of independent and identically distributed random variables extracted from a continuous probability density, is  $P(k) \sim \exp \left(- \gamma _{\text{un}}k \right)$ with $\gamma _{\text{un}} = \ln (3/2)\approx 0.405$. When the results move away from this critical value $\gamma _{\text{un}}$ we are in the presence of correlations and this value is a border between correlated and chaotic stochastic processes,  i. e. $\gamma < \ln (3/2)$ characterizes a chaotic process whereas $\gamma > \ln (3/2)$ characterizes a correlated stochastic one \cite{lacasa2010time}.

\begin{figure}[ht]
\centering
\includegraphics[width=\linewidth]{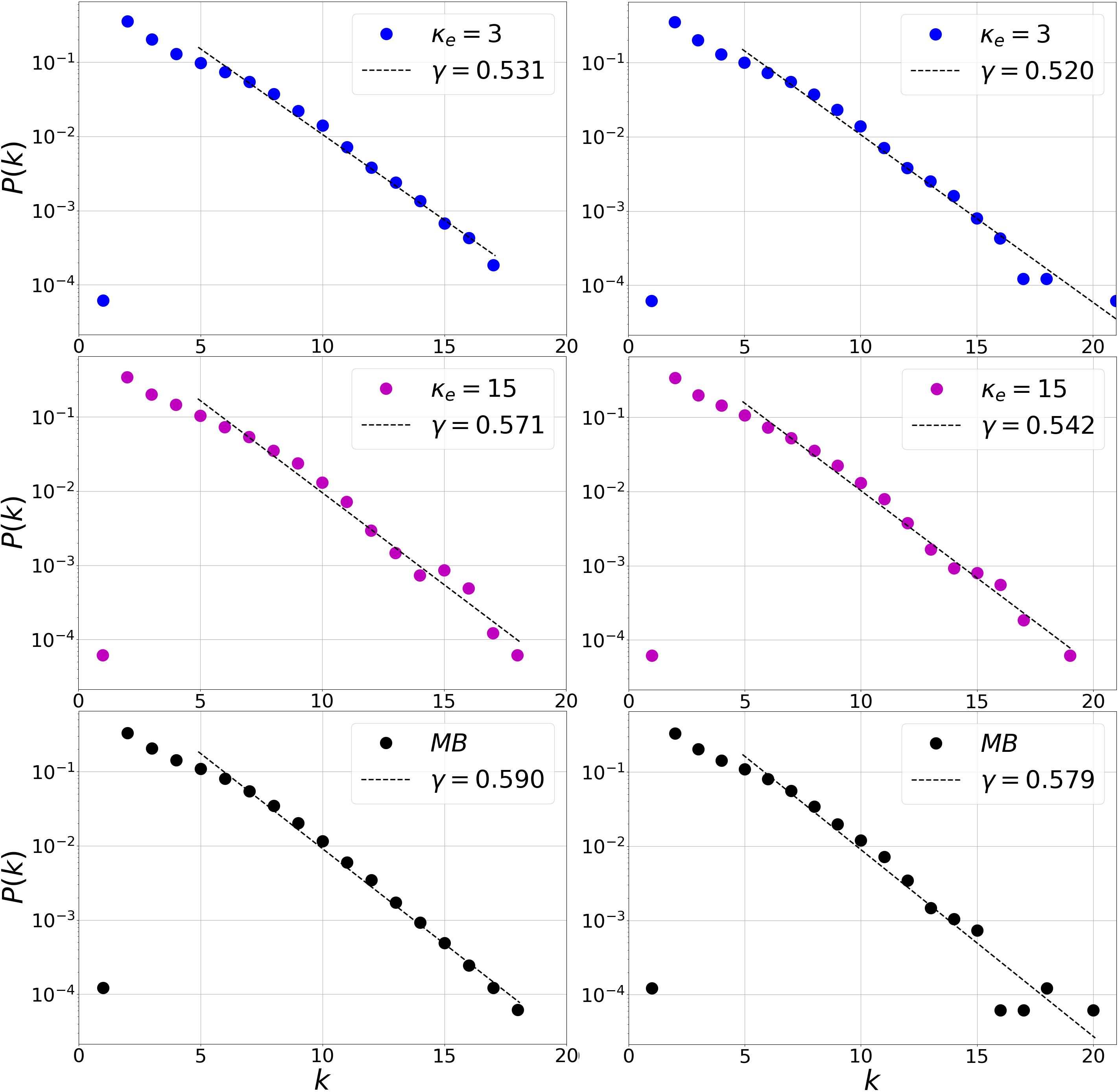}
\caption{Semi-log plot of the degree distributions of HVG associated to Kappa and Maxwell-Boltzmann distribution. There is an exponential behavior $P(k) \sim \exp \left(- \gamma k \right)$ and the $\gamma$ value is shown for each distribution. The left panel corresponds to the results for the magnetic field of the trend data from Figure~\ref{fig:PIC} (left), while the right panel for the detrended data from Figure~\ref{fig:PIC} (right).}
  \label{fig:Pk}
\end{figure}

If we now focus on the undirected HVG, $k(i) = k_{\text{in}}(i) + k_{\text{out}}(i)$, we obtain the undirected degree distribution $P(k)$.  Figures~\ref{fig:Pk} show an exponential distribution for the degree distribution for the three cases studied, this is understood as a short-range exponentially decaying correlations, where $\gamma$ corresponds to the slope of the linear fit in degree distribution semilog. The values of the slope are computed considering the tail of the distribution~\cite{telesca2012} in Figures~\ref{fig:Pk}, in this case from the degree $k =$ 5 up to the largest value of $k$ at each plot. The values of the slope are between $\gamma =$ 0.531 ($\kappa_e =$ 3) and $\gamma =$ 0.590 (Maxwell-Boltzmann distribution), in the case of trended data and between $\gamma =$ 0.520 ($\kappa_e =$ 3) and $\gamma =$ 0.579 (Maxwell-Boltzmann distribution), in the case of detrended data sets. We observe that for each value of the slope it is satisfied that $\gamma > \gamma _{\text{un}}$. This shows us all series corresponds to correlated stochastic processes from which we can extract consistent information. Also, the trend does not seem to greatly affect these correlations.

\begin{figure}[ht]
\widefigure
\includegraphics[width=\linewidth]{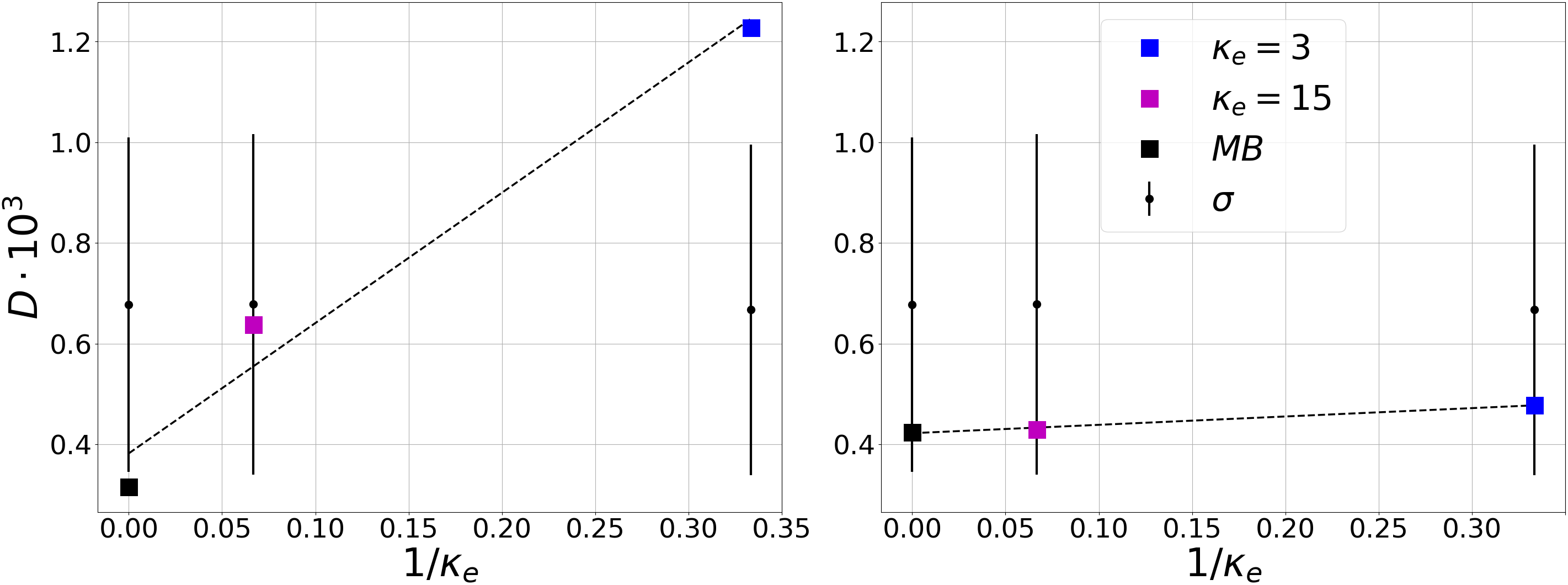}
\caption{KL-Divergence ($D$) of magnetic field for different Kappa distributions. (Left) HVG method applied on the original data. (Right) HVG on the detrended data. The technique used to determine whether the data represent a reversible process consists of applying the HVG algorithm to randomly disordered copies of the data, obtaining the standard deviation $\sigma$ around the average divergence computed using the disordered data (black dot and vertical lines). \label{fig:KLD}}
\end{figure}

Second, considering the directed HVG we have computed the Kullback-Leibler Divergence, $D$ from Equation~\ref{eq:KLD}, for each case mentioned before. The values of the divergence $D$ are in Figure~\ref{fig:KLD} compared to standard deviation $\sigma$ (vertical bars in the figure) calculated from the algorithm to the disarrayed randomly data ~\cite[see e.g.][and references therein]{luctelesca2020time}. In this algorithm the original data in randomly shuffled to obtain a large number of disordered copies (in this case 1000 copies) of the original data set, and the divergence $D$ is computed for each copy. The vertical lines in Figure~\ref{fig:KLD} correspond to the average value of the divergence $D$ of all copies (central value) plus and minus a standard deviation.
After this procedure, if the $D$ value is contained inside the $\sigma$ bar, the time series represents a reversible process. This is because by randomly disarraying the series and obtaining the same results regardless of the temporal order of the data set, by definition it indicates that the information corresponds to a reversible process. On the contrary, if $D$ is outside the vertical range defined by the random copies, then the value of $D$ is statistically significant and therefore it is possible to state that the data set indeed represents an irreversible process.

\begin{figure}[ht]
\widefigure
\includegraphics[width=\linewidth]{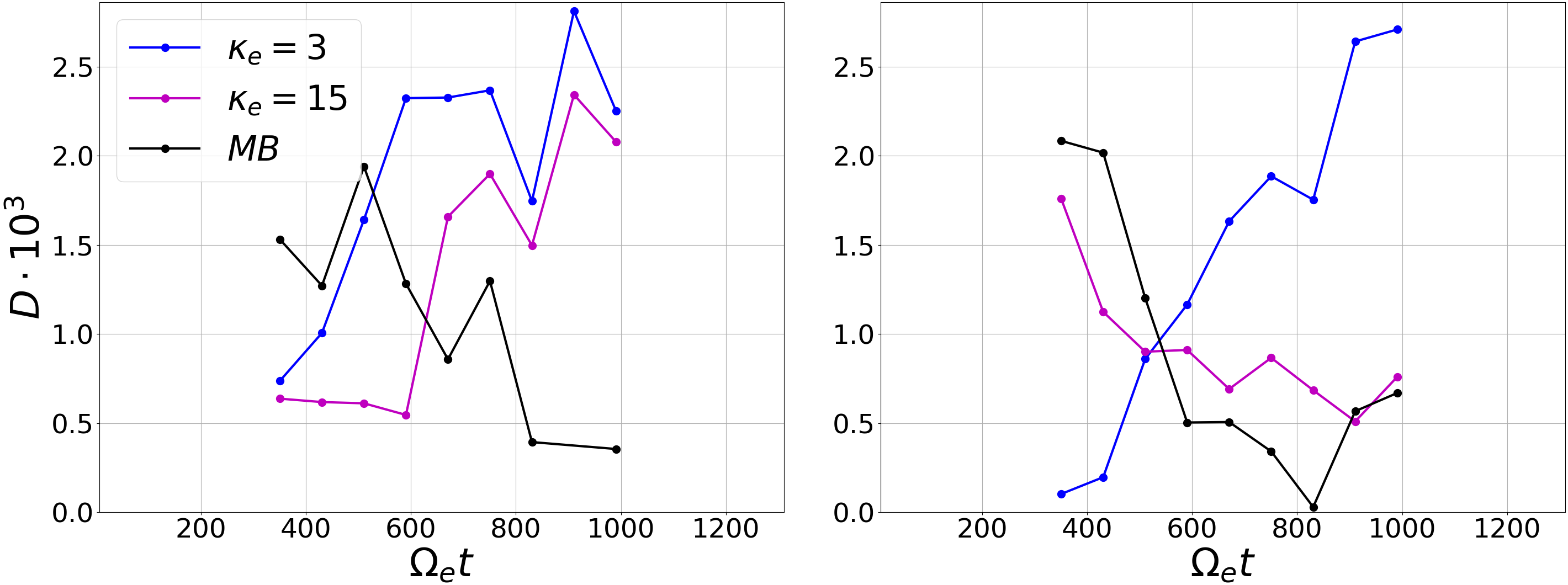}
\caption{Temporal evolution of the KL-divergence considering a moving window that covers $8000$ data overlapping every $1000$ data on the magnetic time series. (Left) HVG method applied on the original data and (right) on the detrended data.\label{fig:KLDtime}}
\end{figure}

Figure~\ref{fig:KLD} shows that the dissipative degree of the system increases as the value of $\kappa_e$ decreases and the distribution function departs from the Maxwell-Boltzmann equilibrium. In Figure~\ref{fig:KLD} (left), the processes for $\kappa_e = 15$ is reversible, case close to thermal equilibrium. Meanwhile, in Figure~\ref{fig:KLD} (right) all distributions correspond to reversible processes by reducing the background trend. This last result could  be explained due to the fact that, independent of the value of $\kappa_e$, all considered distributions are steady state solutions of the Vlasov equation. Finally, to further analyze the relationship between the $\kappa$ parameter and the KLD, we compute the time evolution of $D$ as shown in Figure~\ref{fig:KLDtime}. Figure~\ref{fig:KLDtime} (right) show the same behavior found above, exhibiting a decrease in the value of the divergence for the Maxwellian distribution, whereas for $\kappa_e = 3$ this value tends to increase. That is, given the initial conditions of the simulation, $\kappa_e = 15$ and Maxwellian coincide in their behavior over time towards a low degree of divergence, while $\kappa_e = 3$ evidently presents a behavior to the opposite extreme.\\

\section{Discussion and Conclusions}
\label{Discussion}

In this study we have modeled a turbulent plasma as a complex network, applying the method known as Horizontal Visibility Graph to study the reversibility on magnetic fluctuations. We have developed algorithms to build HVGs from magnetic field fluctuations time series obtained from PIC simulations of collisionless magnetized plasmas. We have analyzed three cases for the time series: a time series of a plasma far from the thermodynamic equilibrium ($\kappa_e =$ 3), a time series closer to the thermodynamic equilibrium ($\kappa_e =$ 15) and a Maxwell-Boltzmann distribution, representing a plasma in a thermal equilibrium. 
For these three time series we have computed the degree distribution of the connectivity, that gives information associated with the time correlations in the distribution and the KLD, that provides information related to the reversibility of the time series. 

In the case of the degree probability distribution we have found an exponential behavior for all cases analyzed, i.e, a short-range correlations for all time series (Kappa and Maxwell-Boltzmann distributions). Our results show that the decaying critical exponent $\gamma$ is the largest for the Maxwellian-Boltzmann distribution, and decreases with decreasing kappa value. Moreover, for $\kappa_e =$ 3 the critical exponent is closer to the limit value $\gamma_{\text{un}}=\ln(3/2)$ proposed by~\citet{lacasa2010time} in which the time series becomes uncorrelated, being chaotic for smaller values ($\gamma < \gamma_{\text{un}}$). These results suggest a lower time correlation for $\kappa_e =$ 3 than the Maxwell-Boltzmann distribution, which is consistent with the fact that in collisionless plasmas out of thermodynamic equilibrium long-range interactions dominate~\cite{Yoon2019}. As already mentioned, in all cases our simulations correspond to isotropic plasmas that are steady-state solutions of the Vlasov equation. Thus, the electromagnetic fluctuations correspond to spontaneous emissions of a system composed by discrete charged particles in random motion. Consequently, the fluctuations provide information about the smallest scales where fast short-range interaction dominate. In the case of a plasma system these scales are strongly related with the Debye length $\lambda_D$.

Inside the Debye sphere (a sphere of radius $\lambda_D$) particles interact individually, and outside the Debye sphere long-range collective interactions dominate. This is directly related to the correlations between the particles that produce the magnetic fluctuations, which depend on the shape of the velocity distribution function. In the case of Kappa distributions the Debye length of the plasma is a decreasing function of $\kappa$ that collapses to zero for $\kappa=3/2$~\cite{bryant1996debye}. Therefore, in plasmas described by a Kappa VDF the short-range correlations are less effective since the Debye length is smaller. Outside the Debye sphere the thermal energy dominates the potential energy and the correlations are practically dissolved~\cite{Livadiotis2018}. In contrast, since the Debye length is greater, in a Maxwellian plasma the short-range correlations dominate, as they decay faster, both temporally and spatially. Regarding our results, this is reflected in the gamma value that seems to behave as a increasing function of $\kappa_e$. 

In the case of the KLD, for both the original and detrended time series, we have obtained low values of the divergence $D$ for all cases, which is consistent with plasmas in steady state according to the Vlasov equation. However, the method have shown to be sensitive enough to distinguish higher values of irreversibility for the Kappa distribution than the Maxwell-Boltzmann case. The irreversibility associated to the Kappa distributions is related to the non-extensive nature of these distributions~\cite{chame1998}, showing an increase in the value of the KLD for decreasing values of $\kappa$. The increase in the value of the KLD indicates a larger value of the entropy in the system. For Kappa distributions following the dynamics of a non-collisional plasma, particles lose individuality and interact collectively increasing entropy~\cite{Yoon2019}. On the other hand, the Maxwell-Boltzmann distribution shows low values for the KLD, being consistent with the Gibbs-Boltzmann entropy. The Maxwell-Boltzmann distribution is related to low values of entropy, in contrast to non-thermal Kappa distributions where it is possible to find a higher (non-extensive) entropy, associated to electromagnetic long-range interactions that dominate the dynamics in the plasma. 

In summary, considering only the limited information provided by the time series, our results seem to indicate a robust relation between the shape of the VDF (given by the Debye length and its dependence on $\kappa$) and the nature of the correlations dominating the magnetic field fluctuations time series represented by $\gamma$~\cite{lacasa2010time}. The connectivity probability distribution shows how the Kappa distribution for low values of $\kappa$ tends to be an uncorrelated time series, while the Maxwell-Boltzmann distribution shows a stochastic time series behavior. Furthermore, we can see that the KLD associated to the HVG is able to distinguish the level of reversibility in time series obtained from PIC simulations and this reversibility seems to be associated to the thermal equilibrium in the plasma. Our results suggest a high sensitivity of the HVG algorithm and a relationship between KLD, $\kappa$ and the entropy of the system. The technique applied here has allowed us to address the role of non-thermal particles distributions in poorly collisional plasma environments. We expect all these features to provide a framework in which complex networks analysis may be used as a relevant tool to characterize turbulent plasma systems, and also as a proxy to identify the nature of electron populations in space plasmas at locations where direct in-situ measurements of particle fluxes are not available.



\authorcontributions{All authors contributed equally to this work. All authors have read and agreed to the published version of the manuscript.}

\funding{This research was funded by ANID Chile, through Fondecyt grant number 1191351.}


\institutionalreview{Not applicable.}

\informedconsent{Not applicable.}

\dataavailability{The data presented in this study are openly available in Zenodo at \href{https://doi.org/10.5281/zenodo.4624381}{https://doi.org/10.5281/zenodo.4624381}, reference number \cite{belen_acosta_2021_4624381}.}

\acknowledgments{We would like to thank Dr. Hocine Cherifi and the organizers of the "Complex Networks 2020" Online Conference, for a great and fruitful scientific meeting, in which an early version of this work was presented. }

\conflictsofinterest{The authors declare no conflict of interest.} 

\end{paracol}

\reftitle{References}



%


\end{document}